\documentclass[11pt]{article}

\usepackage[utf8]{inputenc}
\usepackage[T1]{fontenc}
\usepackage{lmodern}
\usepackage[margin=1in]{geometry}
\usepackage{graphicx}
\usepackage{amsmath,amssymb}
\usepackage{booktabs}
\usepackage{array}
\usepackage{tabularx}
\usepackage{caption}
\usepackage{float}
\usepackage[hidelinks]{hyperref}
\usepackage{url}

\newcolumntype{Y}{>{\raggedright\arraybackslash}X}

\title{Structural Regularities of Cinema SDR-to-HDR Mapping in a Controlled Mastering Workflow: A Pixel-wise Case Study on ASC StEM2}
\author{
Xin Zhang$^{1}$ \and Xiaoyi Chen$^{1}$\\[0.5em]
\normalsize $^{1}$China Research Institute of Film Science \& Technology\\
\normalsize (Test Institute of Film Technical Quality), Beijing 100086, China
}
\date{}

\begin{document}

\maketitle

\begin{abstract}
We present an empirical case study of cinema SDR-to-HDR mapping using ASC StEM2, a rare common-source dataset containing EXR scene-referred images and matched SDR/HDR cinema release masters from the same ACES-based mastering workflow. Based on pixel-wise statistics over all 18,580 frames of the test film, we construct a three-domain comparison involving EXR source data, SDR release masters, and HDR release masters to characterize their luminance and color structural relationships within this controlled workflow. In the luminance dimension, SDR and HDR masters exhibit a highly stable global monotonic correspondence, with geometric structure remaining largely consistent overall; sparse and structured deviations appear in self-luminous highlights and specific material regions. In the color dimension, the two masters remain largely consistent in hue, with saturation exhibiting a redistribution pattern of shadow suppression, midtone expansion, and highlight convergence. Using EXR as a scene-referred anchor, we further define a pixel-level decision map that operationally separates EXR-closer recovery regions from content-adaptive adjustment regions. Under this operational definition, 82.4\% of sampled image regions are classified as EXR-closer recovery, while the remainder require localized adaptive adjustment. Rather than claiming a universal law for all cinema mastering pipelines, the study provides an interpretable quantitative baseline for structure-aware SDR-to-HDR analysis and for designing learning-based models under shared-source mastering conditions.
\end{abstract}

\noindent\textbf{Keywords:} SDR-to-HDR mapping; HDR imaging; computer vision; image structure analysis; inverse tone mapping; interpretable analysis

\section{Introduction}

SDR-to-HDR conversion is an important problem in image and video processing, with applications in display adaptation, content remastering, and visual enhancement. Alongside the deployment of cinema HDR systems and broader HDR standardization efforts~\cite{dci_hdr_addendum,iso_22028_5,iso_21496_1,meininger2019}, existing inverse tone-mapping (ITM) methods mainly focus on perceptual enhancement or scene-referred reconstruction from SDR signals~\cite{cyriac2020,masia2017,itur2446}. However, the structural relationship between professionally mastered cinema SDR and HDR image pairs in real production workflows remains insufficiently characterized.

With the improvement of projection hardware performance and the gradual popularization of HDR technology, there is a substantial real-world demand for adapting a large archive of SDR films to HDR versions. Current mainstream ITM methods attempt to invert the physical world's luminance distribution from a single SDR signal~\cite{banterle2017}. The implicit assumption of such methods is that SDR can be regarded as a compressed record of a physical scene that can be physically recovered through algorithmic means~\cite{itur2446}. However, a cinema release master is a comprehensive product of multiple production decisions including exposure control, composition selection, color grading stylization, and narrative expression. Its tonal and color structure has already been intentionally solidified at the mastering stage, with the goal of serving the narrative structure and visual expression. Therefore, cinema SDR-to-HDR adaptation is better understood as a structured remapping process under different display conditions than as a direct physical inversion of SDR signals.

Existing research has not yet characterized the structural relationship between SDR and HDR release masters in the context of real film production workflows. There is also a lack of systematic quantitative evidence for whether a stable mapping correspondence exists between the two, or for the statistical properties of their differences. These gaps make it difficult to distill universal structural laws and build reliable mathematical models needed for effective SDR-to-HDR conversion.

The American Society of Cinematographers (ASC) Standard Evaluation Material II (StEM2), as an authoritative test film for next-generation digital cinema systems, provides exceptional data conditions for studying SDR-to-HDR mapping~\cite{dci_hdr_addendum,stem2}. The test film provides not only the DCP release masters of SDR (DCI-P3, Gamma 2.6, 48 cd/m$^2$, 12-bit JPEG 2000) and HDR (DCI-P3, PQ, 300 cd/m$^2$, 12-bit JPEG 2000), but also the EXR source data in ACES AP0 linear space (16-bit float). Since the three representations originate from the same mastering pipeline, they provide a rare common-source basis for analyzing SDR-to-HDR structural regularities. Because the present study is based on a single but uniquely well-controlled test film, our goal is not to claim universal laws across all cinema workflows, but to establish a measurable baseline in a shared-source mastering setting that is rarely accessible in public materials.

The main contributions of this paper are:
\begin{enumerate}
\item A three-domain comparative framework based on EXR, SDR, and HDR as co-registered references within a shared ACES-based mastering workflow.
\item Pixel-wise evidence, within StEM2, of highly stable global monotonic luminance mapping across all 18,580 frames.
\item Identification of sparse and structured local residual patterns and their physically motivated interpretation.
\item An operational pixel-level decision map that compares EXR proximity to separate EXR-closer recovery regions from content-adaptive adjustment regions.
\end{enumerate}

\section{Data and Methods}

\subsection{Experimental Data}

ASC StEM2 ``The Mission'' is a comprehensive test film designed around explicit system-testing objectives using a reverse-design methodology. Its script structure, shooting scheme, and post-production workflow were specifically designed to test next-generation digital cinema systems, with content covering dynamic range boundaries, wide color gamut coverage, compression encoding robustness, motion artifact control, and system consistency in LED Volume virtual production.

In the content design phase, the film introduced various scenes with high contrast and extreme lighting conditions. For example, the opening cave segment constructs a high-contrast environment with large dark areas and bright point light sources coexisting via LED virtual production; the car-interior driving segment simulates extreme contrast ratios using heavily overexposed backgrounds; the color of special-effects crystal props is pushed to the Rec.\ 2020 boundary to examine the saturation limit under wide color gamut; the film also contains fast panning shots (for motion artifact testing) and large smoke-filled gradients (for quantization banding testing). These designs make the film an effective carrier for HDR system capability testing.

\begin{figure}[t]
\centering
\includegraphics[width=0.95\linewidth]{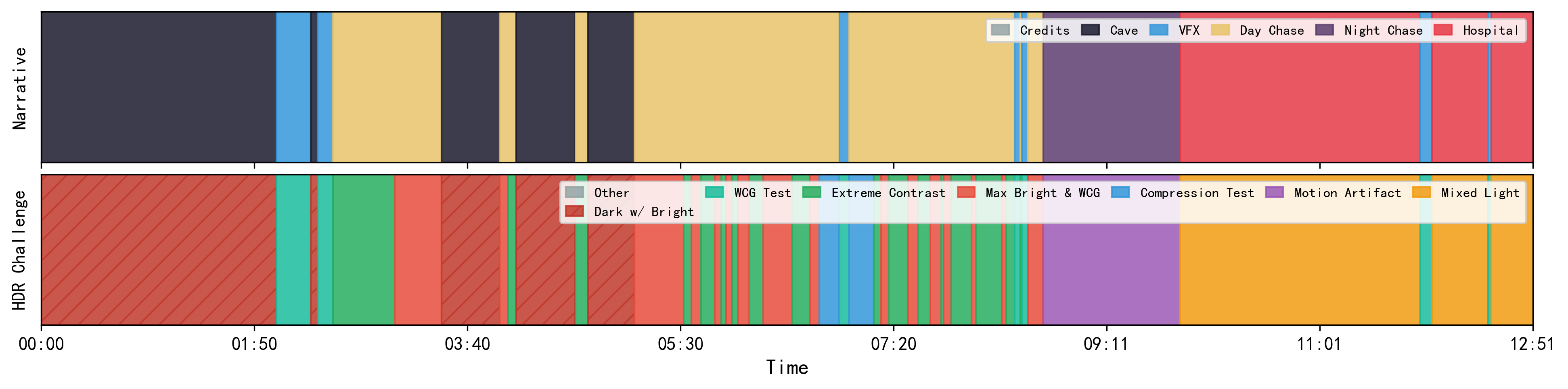}
\caption{Timeline of StEM2.}
\end{figure}

At the production workflow level, StEM2 adopts the Academy Color Encoding System (ACES) as its full-pipeline color management framework~\cite{smpte2065}. The original footage and virtual assets are represented in ACES AP0 linear space, from which SDR and HDR release masters are generated through different Output Device Transforms (ODT). Therefore, the SDR and HDR versions are not two independently authored outputs, but two display-targeted expressions derived from a shared scene structure. This common-source characteristic means that differences among EXR source data, SDR master, and HDR master are primarily determined by the display target and perceptual adaptation strategy. The HDR master's electro-optical conversion uses the ST 2084 Perceptual Quantization curve (PQ EOTF) to achieve a closer match with human visual perception~\cite{smpte2084}.

\subsection{Data Domain Definition}

To avoid confusion between different data attributes, this paper classifies the above data into two conceptually distinct domains.

\begin{enumerate}
\item \textbf{Scene-Referred Domain}. The scene-referred domain is represented by EXR linear data, whose values are approximately linearly related to scene radiance, with an extremely wide dynamic range and color gamut coverage. Data in this domain has not undergone perceptual adaptation for a specific display device, nor has it introduced color grading decisions, thus more closely resembling the original record from the shooting phase.
\item \textbf{Display-Referred Domain}. The display-referred domain is represented by the final distributed SDR and HDR versions, shaped by mastering decisions and target viewing conditions.
\end{enumerate}

\subsection{Analysis Workflow and Statistical Methods}

This study performs a full statistical analysis on all 18,580 frames of StEM2 by frame-by-frame scanning, supplemented by scene-based random sampling covering various typical shooting conditions including day, night, interior, vehicle interior, and VFX scenes. For clarity, the analysis is reported at three levels: full-frame pixel statistics over 18,580 frames, shot-level statistics over 204 sampled shots, and EXR-referenced decision-map analysis over 91 representative frames. Non-narrative segments were excluded from the main statistics; representative simple-graphic frames were examined separately as negative-control illustrations.

\paragraph{Data preprocessing and correspondence.}
Because the EXR source, SDR master, and HDR master are distributed from the same mastered timeline, pixel correspondence is established by identical frame indices and native raster coordinates; no optical-flow-based registration or geometric warping is applied. Before metric computation, we verify that active image areas and letterbox regions are consistent across the three representations, and comparisons are restricted to the common active picture region when necessary.

\paragraph{Absolute-luminance decoding and color transforms.}
SDR release masters are decoded under the published StEM2 cinema mastering condition of DCI-P3 / Gamma 2.6 / 48 cd/m$^2$, while HDR masters are decoded under DCI-P3 / PQ / 300 cd/m$^2$~\cite{dci_hdr_addendum,stem2,smpte2084}. Thus, $L_S$ and $L_H$ in the following equations denote display-referred absolute luminance after transfer-function inversion. EXR data remain in ACES AP0 linear space until the specific stage where XYZ or ICtCp conversion is required.

\paragraph{Sampling and aggregation protocol.}
Full-frame statistics use all frames. Shot-level statistics are obtained by first computing frame-level metrics and then averaging them within each shot; scene-level entries in Tables~1, 5, and 6 are arithmetic means over the sampled shots assigned to that scene category. Representative shots are selected by stratified random sampling across the seven scene types listed in Table~1, and the 91 frames used for decision-map analysis are then sampled from those selected shots with the pseudo-random seed fixed in the analysis scripts for reproducibility.

\paragraph{(1) Log-Luminance Scatter Analysis.}
By plotting the scatter distribution of corresponding pixels between SDR and HDR images in the log-luminance domain, the overall mapping morphology is visually presented, revealing the luminance correspondence between the two.

\paragraph{(2) Isotonic Regression.}
Using the order-preserving regression method to fit the optimal monotonically non-decreasing function~\cite{busing2022}, verifying whether a stable global order-preserving relationship exists between SDR and HDR. Its objective function is:
\begin{equation}
\min_{f} \sum_i (L_H(i) - f(L_S(i)))^2 \quad \text{s.t. } f \text{ is monotonically non-decreasing}
\end{equation}

In Equation~(1), $L_S(i)$ and $L_H(i)$ represent the SDR and HDR luminance values of the $i$-th pixel, respectively. The goodness of fit is quantified by the coefficient of determination ($R^2$) in Equation~(2):
\begin{equation}
R^2 = 1 - \frac{\sum_i (L_H(i) - \hat{f}(L_S(i)))^2}{\sum_i (L_H(i) - \overline{L_H})^2}
\end{equation}

\paragraph{(3) Structural Consistency Analysis.}
To compare the similarity of different versions at the geometric structure level, this paper calculates the Pearson correlation coefficient $\rho$ of images in the gradient domain. The Sobel operator is used to extract gradient magnitudes, and the correlation of corresponding pixels is calculated to evaluate the degree of preservation of edge and texture structures.
\begin{equation}
\rho = \frac{\sum_i (G_S(i) - \bar{G}_S)(G_H(i) - \bar{G}_H)}{\sqrt{\sum_i (G_S(i) - \bar{G}_S)^2} \cdot \sqrt{\sum_i (G_H(i) - \bar{G}_H)^2}}
\end{equation}

In Equation~(3), $G_S(i)$ and $G_H(i)$ are the gradient magnitudes of the SDR and HDR images at pixel $i$, and $\bar{G}_S$, $\bar{G}_H$ are their respective global means. The correlation is computed per frame over all valid pixels and then aggregated to the shot and scene levels as described above. A $\rho$ close to 1 indicates highly consistent geometric structure between the two versions.

\paragraph{(4) Color Space and Color Difference Metrics.}
In color analysis, this paper adopts the ICtCp perceptually uniform color space, using $\Delta E_{ITP}$ as the pixel-level difference metric. This metric offers better perceptual uniformity under HDR and wide color gamut conditions, making it more suitable for structural comparison across display-referred containers. Chroma is defined as $C = \sqrt{C_T^2 + C_P^2}$ and hue angle as $h = \mathrm{atan2}(C_P, C_T)$, so that the reported hue and chroma statistics are directly derived from ICtCp coordinates.

\section{Statistical Structural Relationships of Luminance and Color}

\subsection{Luminance Structural Relationship}

\subsubsection{Global Monotonic Mapping Characteristics}

Figure~2 shows the SDR-to-HDR mapping distribution of all StEM2 pixels in the absolute luminance domain, with a clear monotonic correspondence between the two. The cyan curve represents the global mapping baseline estimated from isotonic regression, and the gray dashed line is the identity mapping reference. It can be observed that the HDR version primarily extends the luminance ceiling in highlight regions. In the low-luminance region, as a reference baseline is introduced for the minimum effective cinema black level, the mapping relationship shows a systematic offset, with the perceptual effect being more gradation detail in the shadows.

\begin{figure}[t]
\centering
\includegraphics[width=0.72\linewidth]{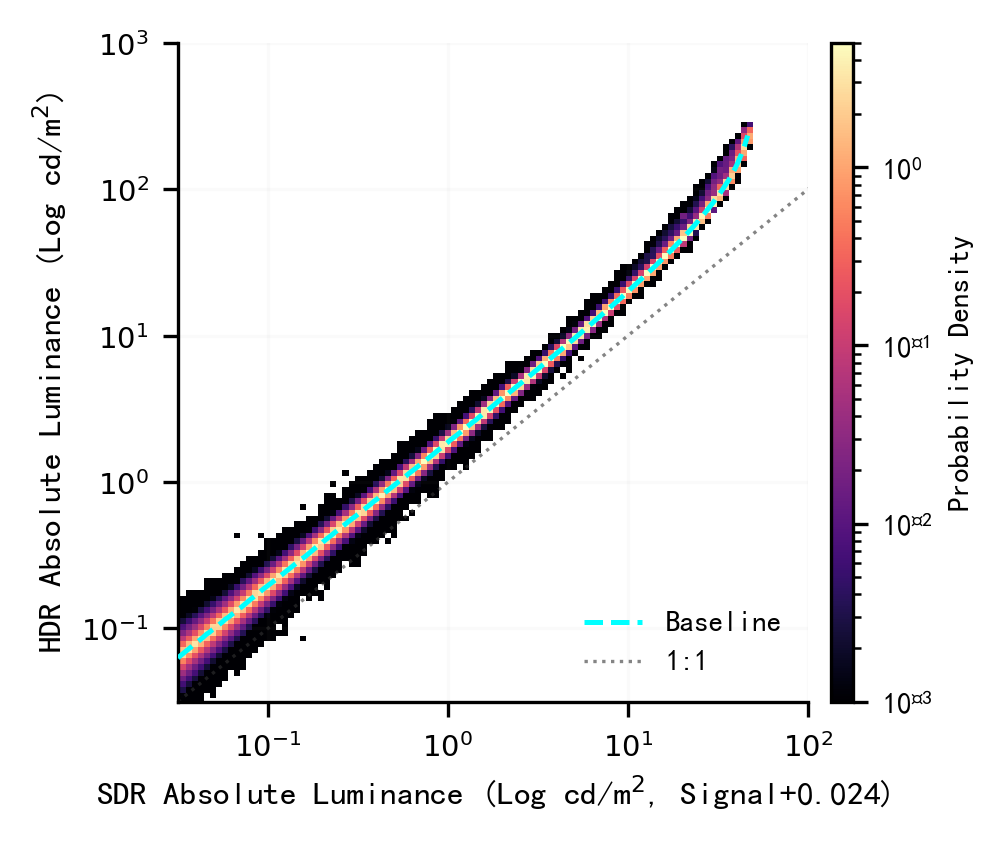}
\caption{Global mapping relationship between SDR and HDR masters in the luminance domain.}
\end{figure}

Statistical results show that for over 85\% of shots, the coefficient of determination ($R^2$) for the SDR-HDR luminance correspondence exceeds 0.99, with a full-film mean of approximately 0.9986. Table~1 presents the statistics classified by scene. In mainstream scenes such as Day Car Interior and Night Interior, the mean $R^2$ reaches above 0.999; even in the Cave scene with the most extreme contrast, the mean $R^2$ still reaches 0.996, with the minimum value of 0.917 occurring in a frame containing intense localized lighting. This suggests that the two masters are not independently authored outputs, but two display-targeted expressions derived from a shared scene structure. Furthermore, this characteristic is consistent with the perceptual monotonicity principle of the Electro-Optical Transfer Function (EOTF) in ITU-R BT.2100~\cite{bt2100}, indicating that HDR dynamic-range expansion is built on strong preservation of the scene's luminance structure. The statistics in Table~1 show that SDR and HDR are highly consistent in edge positions and texture structures, with the gradient correlation coefficient ($\rho$) of the vast majority of shots above 0.96, confirming that HDR production did not introduce significant geometric structural rearrangement.

\begin{table}[t]
\centering
\small
\caption{StEM2 SDR-HDR luminance mapping statistics by scene.}
\begin{tabular}{lcccc}
\toprule
Scene & Number of Shots & Mean ($R^2$) & Minimum ($R^2$) & Gradient Correlation ($\rho$) \\
\midrule
Day Car Interior   & 57 & 0.9995 & 0.9927 & 0.963 \\
Night Car Interior & 21 & 0.9993 & 0.9966 & 0.980 \\
Cave               & 39 & 0.9963 & 0.9169 & 0.944 \\
Desert             & 32 & 0.9990 & 0.9795 & 0.970 \\
Hybrid VFX         & 10 & 0.9986 & 0.9942 & 0.970 \\
Night Interior     & 40 & 0.9995 & 0.9965 & 0.980 \\
Smoke              & 5  & 0.9992 & 0.9964 & 0.959 \\
\bottomrule
\end{tabular}
\end{table}

As shown in Figure~3, under different shot conditions, the monotonic correspondence between SDR and HDR masters in the absolute luminance domain remains stable. For scenes containing strong highlights or self-luminous elements, the luminance distribution shows significant extension in the highlight range; while in scenes with low luminance or well-controlled overall contrast, the distribution is compact and concentrated.

\begin{figure}[t]
\centering
\includegraphics[width=\linewidth]{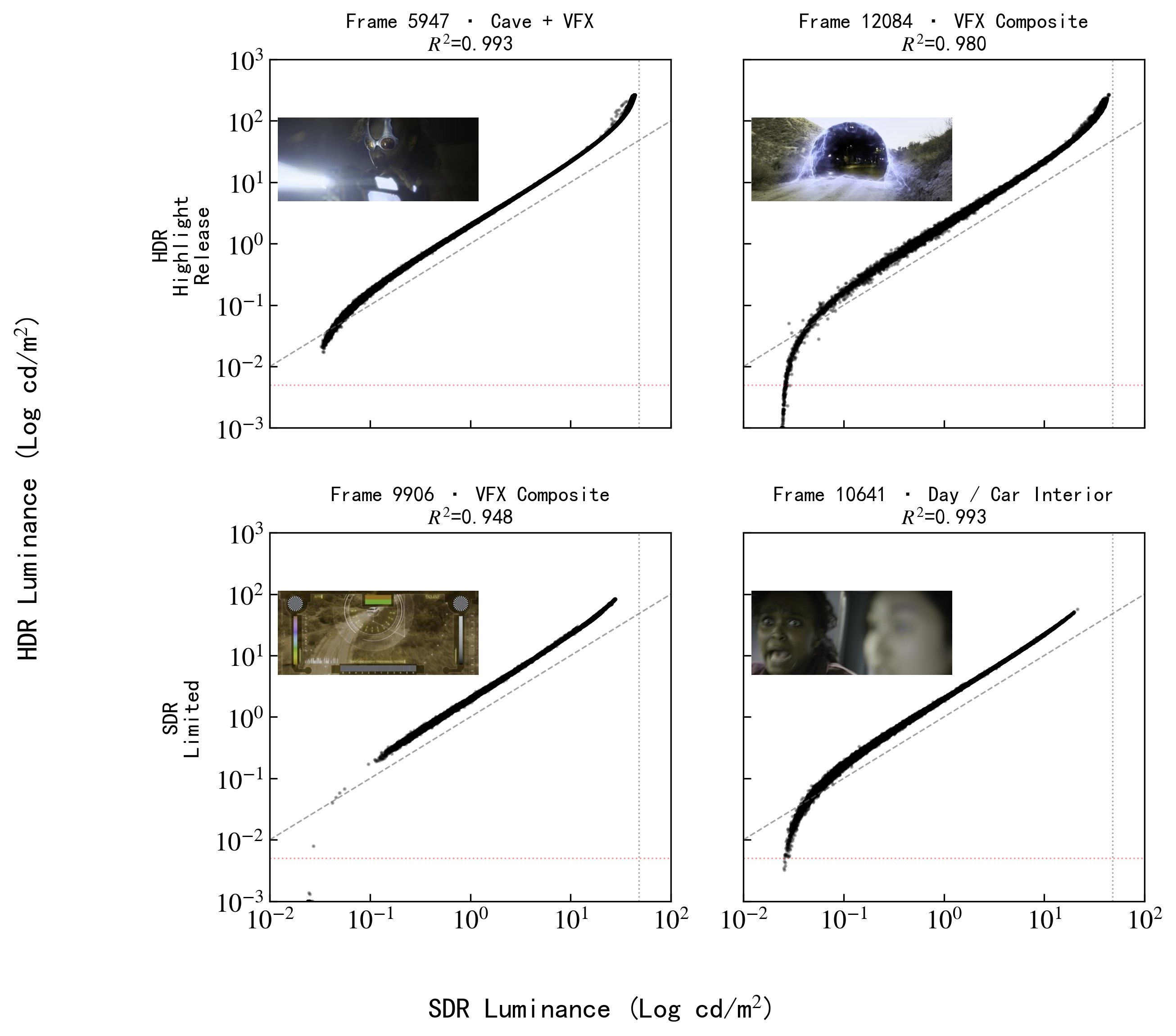}
\caption{SDR-HDR mapping characteristics of typical scenes.}
\end{figure}

\subsubsection{Luminance Residual Structure}

To identify localized adjustments for specific content in HDR production, these deviations need to be decoupled from the global luminance mapping. To this end, we introduce luminance residuals to describe the difference between HDR luminance and the global monotonic baseline prediction. This difference reflects content-dependent local deviations implemented for specific image regions during HDR production, independent of the global strategy.

\begin{figure}[t]
\centering
\includegraphics[width=0.72\linewidth]{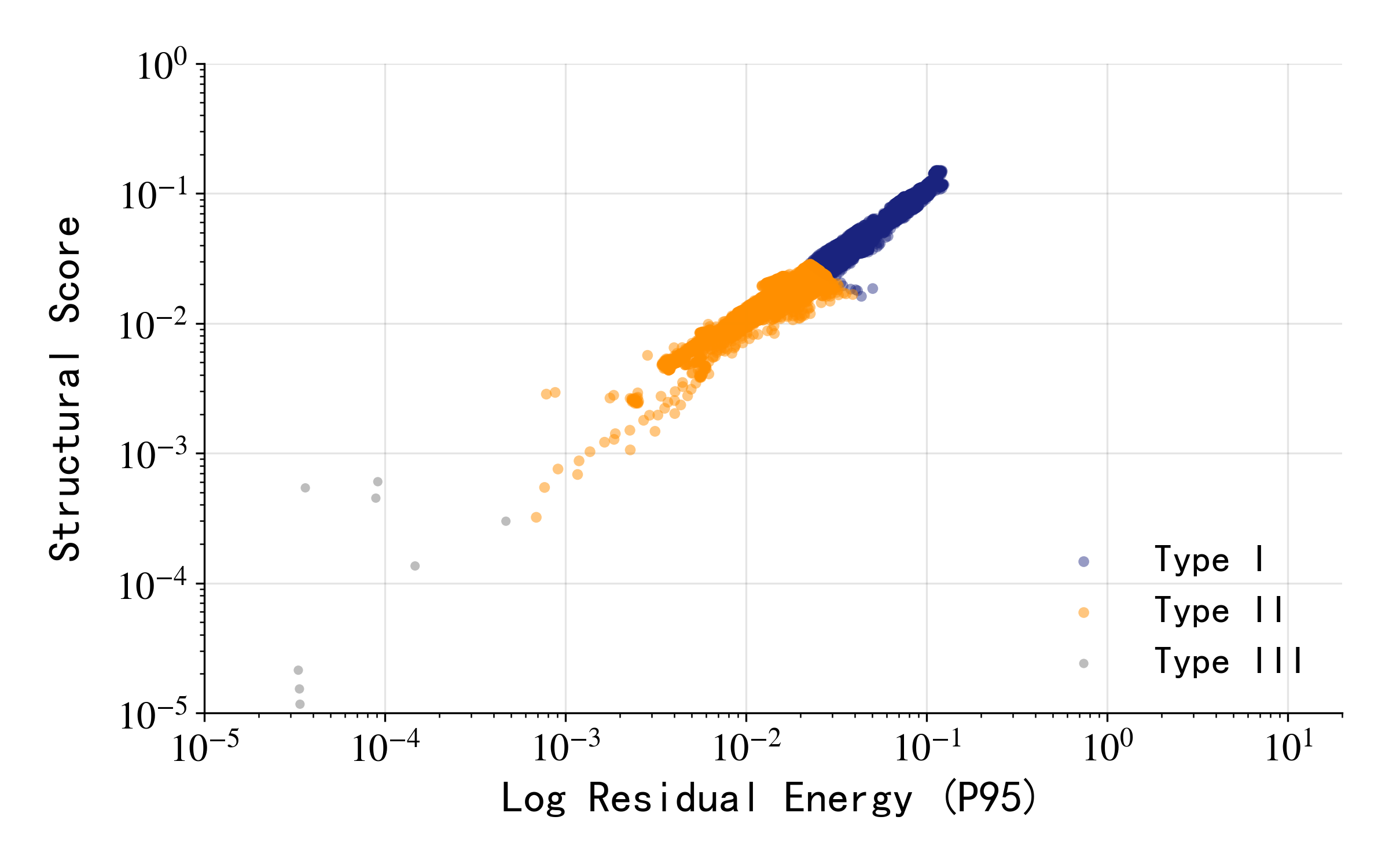}
\caption{Residual clustering in the energy-structure space and physical attribution.}
\end{figure}

\paragraph{(1) Residual Energy and Structure Metrics.}
This paper projects luminance residuals into a 2D ``Energy-Structure'' feature space (see Figure~4). The pixel-level luminance residual is defined as in Equation~(4):
\begin{equation}
\Delta L(i) = L_H(i) - \hat{f}(L_S(i))
\end{equation}

In Equation~(4), $L_S(i)$ and $L_H(i)$ are the SDR and HDR luminance values of pixel $i$, and $\hat{f}(\cdot)$ is the global mapping baseline fitted by isotonic regression (Eq.~1). The residual energy metric uses the 95th percentile of its absolute value $E_{P95} = \mathrm{Percentile}_{95}(|\Delta L|)$, and the structure metric is characterized by the mean gradient magnitude of the residual map $S = \mathrm{mean}(|\nabla \Delta L|_2)$. Before clustering, the 2D feature vector $(E_{P95}, S)$ is z-score normalized across the sampled frames to prevent the energy axis from dominating the partition. K-Means clustering ($k=3$) is then applied to perform unsupervised classification of full-film sample points, resulting in three clusters with distinct physical interpretations.

\paragraph{(2) Three Typical Residual Types.}
\textbf{a. Type I: Self-luminous Highlights.} Type I residuals are mainly distributed in regions with strong self-luminous elements, such as spotlights in the cave scene or sparks in explosion scenes. Their statistical characteristics show significantly elevated residual energy ($E_{P95} \gg 1.0$), with residual edge positions highly coinciding with the contours of luminous bodies. To ensure the overall exposure and visibility of the main subject, the SDR master often clips or compresses highlight signals exceeding display capabilities. In the HDR master, the higher luminance ceiling allows previously constrained highlight extremes to be released, presenting more complete highlight detail. Relevant visual perception studies have shown that highlight regions generally carry greater weight in visual attention allocation~\cite{nemoto2015}, consistent with the concentrated distribution of highlight residuals in this study.

\textbf{b. Type II: Material-related Structural Regions.} Type II residuals are mainly distributed in regions such as transparent HUDs, glass reflections, and metallic highlights. Their statistical characteristics show high structure scores but moderate residual energy. This type of residual can be understood as enhancement of material texture: under the premise of maintaining stable overall luminance relationships, limited adjustments are made to detail regions with clear structural semantics. Such differences that are highly correlated with structure often carry structured content-related information~\cite{junyent2015}.

\textbf{c. Type III: Global-baseline Regions.} Type III regions are areas where the luminance relationship is largely explained by the global monotonic baseline alone, with very low residual energy and structure scores. In the absence of complex narrative needs or material representation goals, no additional compensation components are required. Simple-graphic frames (such as title cards) are representative examples of this behavior and are shown separately as qualitative negative-control illustrations; their statistics are included in Table~2 as part of the full-film pixel distribution.

\begin{table}[t]
\centering
\small
\caption{StEM2 residual classification statistics.}
\begin{tabular}{lccc}
\toprule
Type & Pixel Ratio & Energy Ratio & Physical Mechanism \\
\midrule
Type I: Self-luminous Highlights             & 18.3\% & \textbf{95.4\%} & Self-luminous extremes \\
Type II: Material-related Structural Regions & 31.9\% & 3.8\%            & Material texture \\
Type III: Global-baseline Regions            & 49.9\% & 0.8\%            & General scenes \\
\bottomrule
\end{tabular}
\end{table}

From Table~2: (1) Type III low-residual regions dominate in pixel count at approximately 50\%, with an energy contribution of only about 0.8\%; (2) Type I and Type II together account for approximately 50\% of total pixels, yet contribute over 99\% of residual energy, with Type I self-luminous regions accounting for 95\% of residual energy despite occupying only 18\% of pixels; (3) SDR-to-HDR differences are mainly concentrated in structurally and semantically significant regions (highlights and materials), with the overall residual distribution exhibiting a sparse and structured characteristic.

\subsection{Color Structural Relationship}

In addition to luminance, the differences between SDR and HDR are also reflected in color distribution. Unlike luminance, under the same color gamut conditions, color changes are not directly determined by display system capabilities, but are more related to object properties and creative expression. HDR production generally strives to maintain hue stability~\cite{bt2408}. Therefore, it is necessary to examine whether SDR and HDR remain largely consistent in color structure, and whether their variations exhibit stable statistical characteristics.

To avoid the impact of white point differences on chromaticity distributions, this paper applies a Bradford chromatic adaptation transform under DCI-P3 color gamut conditions to physically align SDR and HDR data at the same white point; subsequently, the data is converted to the ICtCp perceptually uniform space for analysis. The following four metrics are used to describe the color differences between SDR and HDR:
\begin{enumerate}
\item \textbf{Hue Stability:} The mean of the absolute value of the hue angle difference in ICtCp space (Mean$|\Delta h|$), where the value is the circular difference between SDR and HDR hue angles, reflecting the overall level of hue shift.
\item \textbf{Hue Outliers:} The 95th percentile of the absolute value of the hue angle difference, characterizing hue maintenance capability under extreme conditions.
\item \textbf{Chroma Correlation:} The Pearson correlation coefficient between SDR and HDR pixels in terms of Chroma, aiming to evaluate the overall structural consistency of color density.
\item \textbf{Saturation Enhancement Ratio:} For a given luminance interval, the proportion of pixels where HDR saturation exceeds SDR, used to quantify the compensatory release of color volume.
\end{enumerate}

\begin{figure}[t]
\centering
\includegraphics[width=\linewidth]{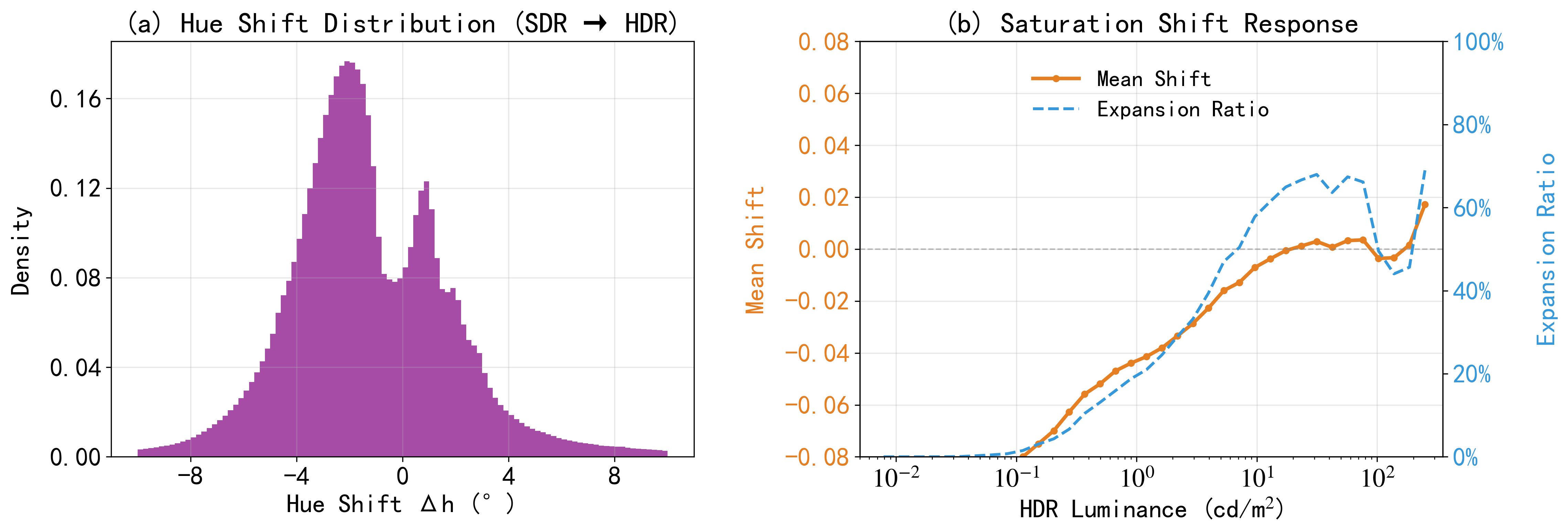}
\caption{Comparison of SDR and HDR color distributions in ICtCp space.}
\end{figure}

\begin{table}[t]
\centering
\small
\caption{StEM2 color statistical metrics (ICtCp space).}
\begin{tabularx}{\linewidth}{l l Y}
\toprule
Metric & Value & Meaning \\
\midrule
Hue Stability (Mean $|\Delta h|$) & 2.38$^\circ$ & Overall hue shift level \\
Hue Outliers (P95 $|\Delta h|$) & 6.52$^\circ$ & Hue maintenance capability under extreme conditions \\
Chroma Correlation & 0.9853 & Degree of chroma correlation between SDR and HDR \\
Saturation Enhancement Ratio (20--100 cd/m$^2$) & 66.9\% & Proportion of pixels where HDR saturation exceeds SDR in the 20--100 cd/m$^2$ region \\
\bottomrule
\end{tabularx}
\end{table}

\begin{table}[t]
\centering
\small
\caption{StEM2 color metrics by luminance bin (ICtCp space).}
\resizebox{\linewidth}{!}{%
\begin{tabular}{lccccc}
\toprule
Luminance Bin (HDR) & Mean Hue Diff ($|\Delta h|$) & P95 Hue Diff & Mean Chroma Change ($\Delta C$) & Saturation Enhancement Ratio & Pixel Ratio \\
\midrule
$<20$ cd/m$^2$   & 2.5$^\circ$ & 5.8$^\circ$  & $-0.039$ & 30.8\% & 85.8\% \\
20--100 cd/m$^2$ & 3.9$^\circ$ & 9.2$^\circ$  & $+0.003$ & 66.9\% & 11.8\% \\
$>100$ cd/m$^2$  & 5.7$^\circ$ & 12.5$^\circ$ & $-0.008$ & 34.4\% & 2.4\% \\
\bottomrule
\end{tabular}%
}
\end{table}

Synthesizing these results, the SDR-to-HDR color mapping exhibits the following main characteristics.

\paragraph{(1) Hue Shift and Stability.}
In Figure~5(a), hue shifts are mainly concentrated near zero, with the majority of pixels showing small hue variation amplitude. The average hue shift given in Table~3 is 2.38$^\circ$, with P95 at 6.52$^\circ$, both small in perceptual terms under complex viewing conditions. This indicates that the two masters remain largely consistent in hue, and the semantic information carried by color remains stable.

\paragraph{(2) Saturation and Color Volume Changes.}
Figure~5(b) shows the distribution of saturation change with luminance. Combined with Table~4, in the midtone region of 20--100 cd/m$^2$, the mean chroma change is $+0.003$, and approximately 66.9\% of pixels show saturation enhancement. The proportion of enhanced pixels in this interval is high but the average amplitude is small, indicating that HDR performs a limited release of color volume in the midtones. In the shadow region ($<20$ cd/m$^2$), the mean chroma change is $-0.039$, with only 30.8\% of pixels showing enhancement; this region also accounts for 85.8\% of pixels. This indicates that shadow pixels are predominantly desaturated in HDR, mainly related to the reduction of perceived saturation as luminance and contrast are re-expanded. In the highlight region ($>100$ cd/m$^2$), the mean chroma change is $-0.008$, with the saturation enhancement ratio dropping to 34.4\%. As luminance approaches the white point, the available color gamut cross-section shrinks, limiting chroma in highlight regions and causing saturation to fall back.

\subsection{Decision-Map Analysis with EXR as Scene Reference}

The preceding analysis has revealed the structural relationships between SDR and HDR in both luminance and color dimensions. However, comparing only the two release masters cannot determine whether observed differences originate from changes in display containers or from creative reconstruction. To this end, we further introduce EXR source data as a scene-referred reference, constructing a three-domain comparison of ``EXR--SDR--HDR'' to derive an operational interpretation of HDR mastering behavior.

\paragraph{(1) Selection of Discriminative Metrics and Measurement Rationale.}
Because luminance changes between different display containers are unavoidable and significantly influenced by EOTF form, peak luminance ceiling, black level recalibration, and visual adaptation, color relationships---especially hue and local contrast structure---exhibit stronger stability in cinema narrative. They are often directly related to object identity, material characteristics, and scene semantics, making them more suitable as proxy indicators for whether creative intent has been preserved. Based on this consideration, we select $\Delta E_{ITP}$ in the perceptually uniform ICtCp space as the measure for classifying HDR mastering behavior.

Since EXR data is a scene-referred relative radiance record while release masters are display-referred absolute luminance values, an unknown global gain coefficient exists between them. To eliminate this difference, EXR (ACES AP0) is first converted to XYZ (D65) space, and a gain compensation is applied to EXR using the mid-gray luminance interval as an anchor, aligning it with the midtone regions of SDR and HDR. After completing this alignment, for each pixel position, the perceptual distance of its SDR and HDR versions relative to the EXR reference is calculated:
\begin{equation}
\Delta E_{SDR}(x) = \Delta E_{ITP}(\text{SDR}(x),\ \text{EXR}(x)), \quad
\Delta E_{HDR}(x) = \Delta E_{ITP}(\text{HDR}(x),\ \text{EXR}(x))
\end{equation}

$\Delta E$ is the structural deviation in perceptual space, always a non-negative quantity used to express proximity. A decision rule is further constructed: if $\Delta E_{HDR}(x) < \Delta E_{SDR}(x)$, the pixel is labeled as an EXR-closer recovery region under our operational definition; if $\Delta E_{HDR}(x) > \Delta E_{SDR}(x)$, it is labeled as a content-adaptive adjustment region. To filter out the influence of texture masking and quantization noise, 3 JND ($\Delta E_{ITP} \approx 3$) is used as an engineering robustness threshold when interpreting small local differences. The resulting pixel-level Decision Map should therefore be understood as an analysis tool referenced to the aligned EXR anchor, rather than as a direct measurement of authorial intent or physical ground truth.

\paragraph{(2) Statistical Distribution Characteristics of the Decision Map.}
Applying the above method to multiple typical scenes of StEM2, the resulting Decision Maps exhibit a binary stratified structure under different lighting conditions and content types, as shown in Figure~6. Green pixels indicate EXR-closer recovery under the operational rule; red pixels indicate content-adaptive adjustment.

\begin{figure}[t]
\centering
\includegraphics[width=\linewidth]{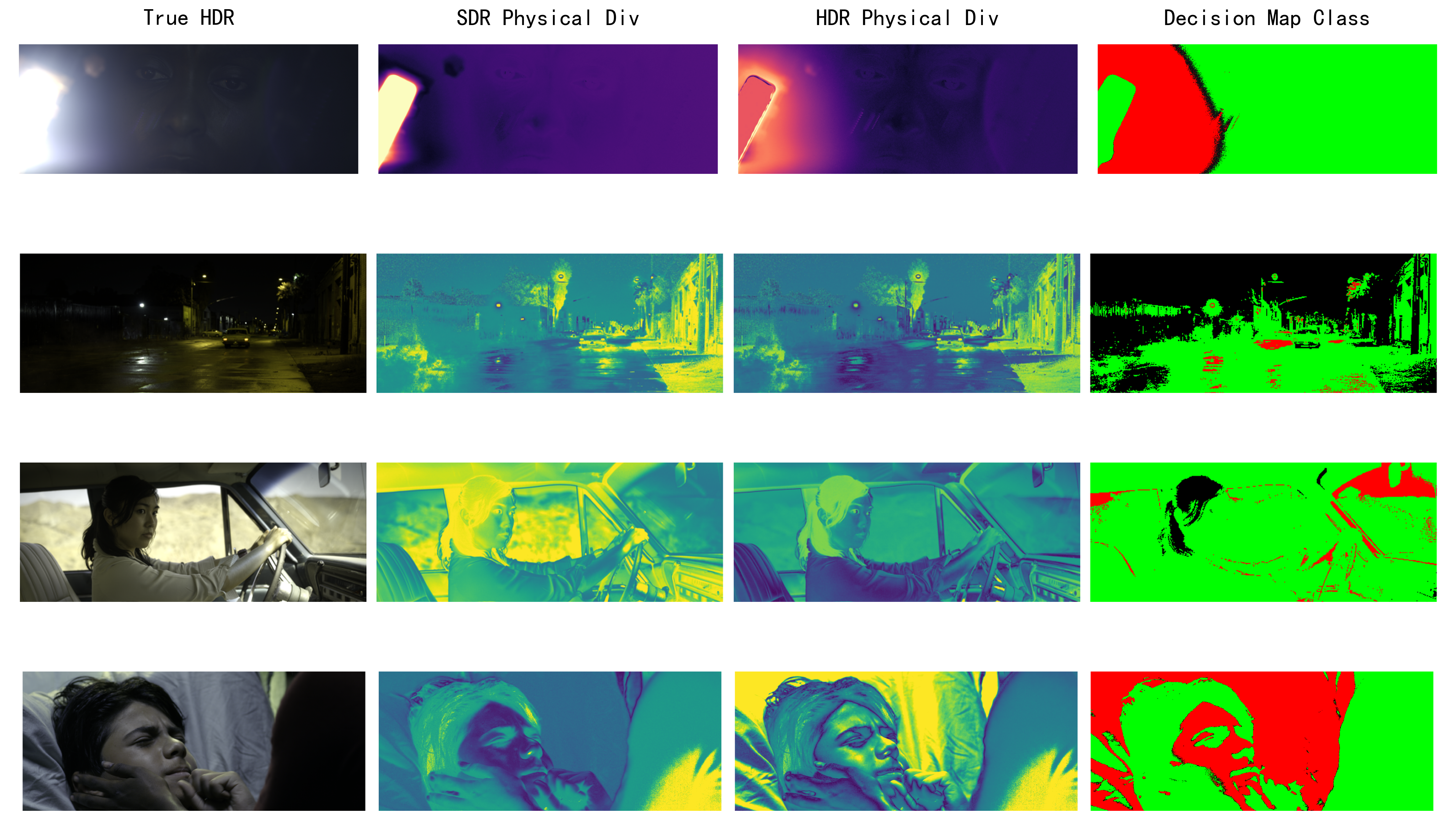}
\caption{Pixel-level decision maps in representative scenes. Green indicates EXR-closer recovery under the operational rule; red indicates content-adaptive adjustment.}
\end{figure}

Statistical results show that EXR-closer recovery regions are statistically dominant. Sampling and analyzing 91 frames from the full film, the proportion of EXR-closer recovery regions is 82.4\% (see Table~5). This indicates that, under the adopted alignment and metric, HDR is more often closer to EXR than SDR in the majority of sampled regions.

\begin{table}[t]
\centering
\small
\caption{StEM2 EXR-closer recovery vs.\ content-adaptive adjustment ratios.}
\begin{tabular}{lcc}
\toprule
Scene & EXR-Closer Recovery (Green) & Content-Adaptive Adjustment (Red) \\
\midrule
Day Car Interior   & 80.0\%  & 20.0\% \\
Night Car Interior & 76.7\%  & 23.3\% \\
Cave               & 77.7\%  & 22.3\% \\
Desert             & 97.9\%  & 2.1\% \\
Hybrid VFX         & 84.4\%  & 15.6\% \\
Night Interior     & 79.6\%  & 20.4\% \\
Smoke              & 100.0\% & 0.0\% \\
\textbf{Full Film Average} & \textbf{82.4\%} & \textbf{17.6\%} \\
\bottomrule
\end{tabular}
\end{table}

In contrast, content-adaptive adjustment regions are mainly concentrated in extreme highlights (such as spotlight centers), high-saturation luminous bodies (such as neon tubes), and specular reflections of specific materials. These regions often correspond to the most significant display constraints, luminance clipping, or color volume compression conditions.

\paragraph{(3) Interpretation of the Operational Classification and Its Implications for SDR-to-HDR Mapping.}
The preceding classification is defined operationally through perceptual color difference metrics, but its likely causes can still be interpreted in combination with luminance structural characteristics and signal-to-noise ratio conditions.

In EXR-closer recovery regions, the HDR version statistically exhibits luminance and color structures that are closer to EXR under the adopted metric. In these regions, the differences between SDR and EXR are mainly manifested as dynamic range compression: information in SDR does not show significant structural destruction but is mainly constrained by the luminance container. These regions therefore provide stronger evidence for physically constrained recovery behavior. However, the degree of recoverability is significantly influenced by the signal-to-noise ratio (SNR) of the original material (Table~6): in high-luminance, high-SNR scenes (such as the Desert), the structural correlation between HDR and EXR is markedly higher than in low-luminance, low-SNR scenes (such as the Cave). In such low-SNR shadow regions, valid detail is often mixed with random noise or has already been lost during compression encoding. These observations suggest that low-SNR shadow regions are less suitable for literal recovery. In such regions, perceptually guided reconstruction may be more appropriate than direct physical inversion.

\begin{table}[t]
\centering
\small
\caption{Structural correlation between StEM2 scene reference (EXR) and release masters (SDR/HDR).}
\begin{tabularx}{\linewidth}{lccY}
\toprule
Scene & Corr.\ (EXR--HDR) & Corr.\ (HDR--SDR) & Notes \\
\midrule
Desert   & 0.72 & 0.993 & High dynamic range, high-SNR scene; SDR mainly constrained by luminance container compression \\
Cave     & 0.16 & 0.888 & Low-luminance, low-SNR scene; structural correlation significantly reduced \\
Hospital & 0.85 & 0.986 & Diffuse-reflection dominated scene; master and scene reference have higher structural consistency \\
\bottomrule
\end{tabularx}
\end{table}

In content-adaptive adjustment regions, residuals are mainly concentrated in self-luminous highlights, specular highlights, and specific material areas. In SDR, these regions often experience severe luminance clipping, saturation compression, or quantization distortion, resulting in the original structure no longer being reliably recoverable from SDR. Therefore, changes in these regions cannot be reversed at the structural level, but luminance and saturation can be compensated while maintaining hue and object identity stability.

\section{Conclusion}

This paper takes ASC StEM2 as the research object and quantitatively characterizes the mapping relationships among EXR source data, SDR, and HDR release masters in luminance and color dimensions through empirical and pixel-wise statistical analysis. Within this controlled mastering workflow, the results show that a stable global monotonic correspondence exists between SDR and HDR masters, with their image structures maintaining a high degree of overall consistency; meanwhile, the differences between EXR and the release masters indicate that tonal and color decisions are mainly solidified at the mastering stage. The Decision Map constructed using aligned EXR as a scene reference further shows that approximately 82.4\% of the sampled regions are labeled as EXR-closer recovery under the adopted operational definition, with content-adaptive adjustments concentrated in self-luminous highlights and specific material regions.

Based on these empirical results, the study supports what we term a restrained restoration view of cinema SDR-to-HDR adaptation: under the premise of maintaining the original narrative structure and perceptual stability, the process selectively releases information constrained by the display container. In the luminance dimension, this process is dominated by a global monotonic baseline, with controlled compensation introduced only in structurally significant regions; in the color dimension, hue is maintained stable, and color volume is compensated within the primary luminance range, while a more conservative mapping strategy is adopted in the extreme highlight range. For learning-based SDR-to-HDR methods, these observations suggest that a strong global monotonic prior can be combined with model capacity concentrated on self-luminous highlights, material-dependent residuals, and low-SNR shadow regions where literal inversion is unreliable. As an engineering reference associated with these observations, we also provide an external implementation resource that integrates SDR-to-HDR processing and HDR DCP packaging in a modified DCP-o-matic workflow~\cite{zhang_dcpomatic}.

Because the evidence reported here is derived from a single controlled test film, the findings should be interpreted as a structured baseline rather than a universal law across all cinema mastering workflows. Extending the analysis to additional titles, mastering teams, and display targets remains necessary before broader generalization.

\end{document}